\documentclass{aa501}
\usepackage{graphicx, lscape}
\begin{document}
   \title{Formation scenarios for the young stellar 
  associations between galactic longitudes $l = 280\degr-360\degr$ 
   \thanks { Tables 1 to 4 are only available in electronic form at the CDS via 
anonymous ftp to cdsarc.u-strasbg.fr (130.79.128.5) or via 
http://cdsweb.u-strasbg.fr/Abstract.html}
              }

   \author{M. J. Sartori
          \inst{1,2}
          \and 
          J. R. D. L\'epine\inst{2}          
          \and 
          W. S. Dias \inst{2}
           }

   \offprints{M. J. Sartori,
   \email{marilia@lna.br}}

   \institute{Laborat\'orio Nacional de Astrof\'{\i}sica/MCT,
			 C.P. 21, 37500-000 Itajub\'a - MG, Brazil\\
         \and
         Instituto de Astronomia, Geof\'{\i}sica e Ci\^encias Atmosf\'ericas,
		 Universidade de S\~ao Paulo, C.P. 3386, 01060-970 
		 S\~ao Paulo - SP, Brazil\\
               }

   \date{Received 12 November 2002 / Accepted 17 March 2003}

   \abstract{
   We investigate the spatial distribution, the space velocities and age 
   distribution of the pre-main sequence (PMS) stars belonging to Ophiuchus, 
   Lupus and Chamaeleon star-forming regions (SFRs), and of the young early-type 
   star members of the Scorpius-Centaurus OB association. These young stellar 
   associations extend over the galactic longitude range from $280\degr$ to 
   $360\degr$, and are at a distance interval of around 100 and 200 pc. 
   This study is based on a compilation of distances, proper motions and 
   radial velocities from the literature for the kinematic properties, and 
   of basic stellar data for the construction of Hertzsprung-Russel diagrams. 
   Although there was no well-known OB association in Chamaeleon, the distances 
   and the proper motions of a group of 21 B- and A-type stars, taken from the 
   Hipparcos Catalogue, lead us to propose that they form a young association.    
   We show that the young early-type stars of the OB associations and the PMS 
   stars of the SFRs follow a similar spatial distribution, i.e., there is no 
   separation between the low and the high-mass young stars. 
   We find no difference in the kinematics nor in the ages of these two
   populations studied. 
   Considering not only the stars selected by kinematic criteria but the whole 
   sample of young early-type stars, the scattering of their proper motions is 
   similar to that of the PMS stars and all the young stars exhibit a common 
   direction of motion.
   The space velocities of the Hipparcos PMS stars of each SFR are compatible 
   with the mean values of the OB associations.
   The PMS stars in each SFR span a wide range of ages (from 1 to 20 Myr).
   The ages of the OB subgroups are 8--10 Myr for Upper Scorpius (US), 
   and 16--20 Myr for Upper Centaurus Lupus (UCL) and for Lower Centaurus Crux (LCC). 
   Thus, our results do not confirm that UCL is older than the LCC association.
   Based on these results and the uncertainties associated with the age determination,
   we cannot say that there is indeed a difference in the age of the two populations.
   We analyze the different scenarios for the triggering of large-scale 
   star-formation that have been proposed up to now, and argue that most probably we 
   are observing a spiral arm that passes close to the Sun. The alignment of young 
   stars and molecular clouds and the average velocity of the stars in the opposite
   direction to the Galactic rotation agree with the expected behavior of star 
   formation in nearby spiral arms. 
   }
\titlerunning{Formation scenarios for young stellar associations}
\authorrunning{M. J. Sartori et al. }
   \maketitle
%

\section{Introduction}
The Ophiuchus, Lupus and Chamaeleon molecular cloud complexes are prominent 
low-mass star-forming regions (SFRs) close to the Sun (100--200 pc), 
situated between galactic longitudes $280\degr$ and $364\degr$, and
reaching relatively high galactic latitudes ($+20\degr$ for $\rho$ Oph and 
$-15\degr$ for Chamaeleon). Their alignment, as can be seen in a projection 
on the Galactic plane (e.g., Dame et al. \cite{dame}), suggest that they are part 
of an elongated structure that extends over more than 120 degrees in longitude, 
from the Aquila Rift to the Vela region. This structure also includes an association 
of massive stars, the Scorpius-Centaurus (Sco-Cen) OB association, with its
subgroups Upper Scorpius (US), Upper Centaurus Lupus (UCL), and Lower 
Centaurus Crux (LCC). 
The prominent molecular cloud complexes (Ophiuchus, Lupus, Chamaeleon) 
are connected by filaments to the Galactic plane, and there seems to be an extended 
thin layer of gas and dust parallel to the aligned structure at distances of about 
100--150 pc from the Sun. This was discovered through studies of reddening as a 
function of distance of stars, which established the physical association between 
the Southern Coalsack and the Chamaeleon and Musca clouds (Corradi et al. \cite{corradi}), 
and from a mapping of Na{\small I} absorption lines (Genova et al. \cite{genova}). 

Our purpose is to investigate the dynamics of the star-formation process
in this extended region of the sky where possibly a dominant mechanism triggered
the formation of all the stars. 
Many different scenarios have been proposed for the formation of this complex 
of stars and clouds, for instance the sequential star-formation process 
proposed by Blaauw (\cite{blaauw64}, \cite{blaauw91}), the Gould Belt model of Olano 
\& Poppel (\cite{olano87}), which considers that a large disturbance originated in 
the Taurus region and is propagating in the Galactic plane, or the direct impact 
of high velocity clouds on the disk proposed by L\'epine \& Duvert (\cite{lepine94}). 

Useful stars to study the dynamics of the star-formation process are those 
that are younger than about 20 Myr. In such a short time, stars cannot move far 
from their birthplace, and their trajectories can be traced back based on their 
present space velocities. Furthermore, the age of groups of stars can be 
estimated from Hertzsprung-Russel diagrams. Two main classes of stars fall in this 
range of ages. 
Most massive stars have already reached the main sequence, or evolved away from 
it, and form the bright OB associations. On the other hand, most stars with mass 
less than a few Solar masses, in the same range of age of the early-type stars
of the OB associations, have not yet reached the main sequence, and are considered 
as pre-main sequence (PMS) stars, and include T Tauri (TT) stars and Herbig Ae/Be 
(HAeBe) stars. A detailed analysis of the motion of the PMS stars and of the  
early-type stars members of OB associations is essential to derive constraints 
to the star-formation models. Different behavior of space velocities, velocity 
gradients and gradients of age of the stellar associations can be predicted from 
the above scenarios and other possible ones. 

Usually the star formation history of these close SFRs and of the Sco-Cen OB 
association has been studied separately for each sub-region, and also separately 
for massive and low-mass stars.
In this work, we compare the spatial distribution and kinematics of the PMS 
stars with those of the young early-type stars. We know that the group of PMS 
associated with the Ophiuchus SFR and the young early-type stars of 
the US OB association are situated in the same region of the sky. Several works
have studied the PMS population of the US association, e.g., Walter et al. 
(\cite{walter}), Mart\'{\i}n (\cite{martin}), Sciortino et al. (\cite{sciortino}), 
Preibisch et al. (\cite{preib98}), Preibisch \& Zinnecker (\cite{preib99}), and 
recently, Preibisch et al. (\cite{preib02}). The PMS group of the Lupus SFR and 
the UCL OB association are also situated in the same region of the sky (Mamajek 
et al. \cite{mamaj02}). Other works, e.g., the PDS survey (Gregorio-Hetem et al. 
\cite{gregorio}; Torres et al. \cite{torr95}; Torres \cite{torr99}), Feigelson 
\& Lawson (\cite{feigel97}), and recently, Mamajek et al. (\cite{mamaj02}), have 
shown that the LCC OB association also contains PMS stars. 
However, there is no well-known OB association in Chamaeleon. Therefore, in this 
work we verify the hypothesis that there also is an OB association in Chamaeleon 
and we verify its relation to the PMS stars. We also re-examine the age 
determination of the different stellar associations. Finally, we discuss the 
consistency of these data with the main star-formation scenarios.


\section{The samples of young stars}

\subsection{Pre-main sequence stars}
Sartori (\cite{sart00}) collected a list as exhaustive as possible of known TT and 
HAeBe stars clearly associated to the SFRs of Ophiuchus, Lupus, Centaurus and 
Chamaeleon, and also those that are situated in an extended region around the SFRs. 
The list contains 610 stars, discovered through X-ray surveys, infrared surveys, 
H$\alpha$ surveys, and other means, that had their nature confirmed by optical 
spectroscopy. Although most TT stars are not bright enough to have been observed by 
the Hipparcos mission, a few of them are found in the Hipparcos catalogue 
(Neuh\"auser \& Brandner \cite{neuhau}; Wichmann et al. \cite{wich98}; Bertout et 
al. \cite{bertout}). In the studied SFRs, 39 of the selected PMS stars are in the 
Hipparcos Catalogue (ESA \cite{esa}). In Table 1 (only available in electronic form 
at the CDS) we present the 570 PMS stars that are not in the Hipparcos Catalogue 
and in Table 2 (only available at the CDS too) the 39 Hipparcos stars. In these 
tables, we present the names of the stars in Col. 1. In Table 2, we also present 
the Hipparcos numbers (Col. 2), the type of the stars (classical or weak-line TT, 
or HAeBe) in Col. 3 and the references of this classification in Col. 4.

\subsection{Young early-type stars}
The sample of young early-type stars is largely based on the work of de Zeeuw et al. 
(\cite{dezeeuw}, hereafter dZ99), who studied the young stars belonging to the 
subgroups of Sco-Cen OB association. In that work, the selection of stars members 
of the associations, based on Hipparcos parallaxes and proper motions, made use of 
a combination of the convergence point method and the ``spaghetti" method of 
Hoogerwerf \& Aguilar (\cite{hooger}). In Table 3 (only available at the CDS), we 
list 521 stars selected by dZ99 as secure members. In this table, the Col. 1 lists 
the HD numbers and Col. 2 the Hipparcos numbers. The secure members selected by dZ99 
are mostly B-, A-, and F-type stars (there are no O-type stars in this OB association), 
but they also selected some G-, K-, and even M-type stars. dZ99 point out that their 
kinematic selection method accepts some field stars as association members and they
describe their method for estimating the expected number of these interlopers. 
Although few G- to M-type stars have been predicted as interlopers, they believe 
that the majority of the G- to M-type members probably do not belong to the 
associations. We discuss the membership of these late-type stars in Sect. 5.   

In their work, dZ99 found a considerable overlap of the secure members with the stars 
they call as classical members (stars traditionally considered as members of the Sco-Cen
OB association). However, they found also significant differences: the vector point 
diagram of the secure members is more concentrated than that of the classical members, 
and the parallax distribution is narrower. They attribute these results to a reduced 
contamination of field stars. Comparing the dZ99 selection with the list of early-type 
stars previously considered as belonging to the Sco-Cen OB association by de Geus et al. 
(\cite{degeus}, hereafter dG89), we verified that several of these classical members 
have been indeed excluded by the method of dZ99. 
We believe, based on the comparison with the PMS star proper motions and on the H-R 
diagrams later discussed (Sects. 4 and 5), that the selection of dZ99 is too restrictive, 
and that many of the stars excluded by these authors are actually members of the 
associations. We decided to also consider in this work these sample of classical members 
excluded by dZ99. We selected in the list of dG89 the stars having Hipparcos parallax 
measurements of good quality, for which the derived distances are closer than 250 pc. 
In this way, we added 35 stars in US, 41 in UCL, and 12 in LCC. The list of these stars 
is presented in Table 4 (only available at the CDS -- Col. 1 lists the HD numbers and 
Col. 2 the Hipparcos numbers).

\subsection{Chamaeleon OB association}
In the Chamaeleon region there was no previously well-known OB association, 
although Eggen (\cite{eggen}) mentioned the existence of an association that could 
be an extension of the Sco-Cen one. We searched in the Hipparcos Catalogue for B- 
and A-type stars situated in the whole region of the sky where Chamaeleon PMS stars 
are found. We selected 21 stars that have distance values in the range of
distances of the PMS stars (between 120 and 220 pc). These stars are presented 
in Table 5, where Col. 1 lists their HD numbers and Col. 2 the Hipparcos numbers, 
Cols. 3--4 give their galactic coordinates and Col. 5 gives the distance of each 
selected star.

We also analyzed the space motions of these stars. Adopting the Hipparcos proper
motions (Cols 9--10 of Table 5) and radial velocities of the literature (values 
in Col. 11 and references in Col. 12), we computed the U, V, W velocity components.
These velocity components, corrected for the Solar motion (see Sect. 4) are
presented in Cols. 13--14 of Table 5. The dispersion of the space velocity 
distribution of these stars (see values in Table 6) suggest that they 
probably belong to a single group. We propose that this group of B- and A-type 
stars form an association, not previously catalogued, in the Chamaeleon region.
We can note (see the values given in Tables 2 and 5) that the space velocities of 
these stars are similar to those of the PMS stars, which reinforces the argument 
that the two populations are related (see discussions in Sect. 4).

The $\eta$ Cha open cluster investigated by Mamajek et al. (\cite{mamaj00}) is contained 
in a field of 40$\arcsec$, which is small compared to the 15$\degr$ size of the
association that we are discussing here. The stars of the $\eta$ Cha cluster
are situated at about 90 pc, closer than the average of the stars of our list,
and the only B-type star of the cluster is $\eta$ Cha itself (spectral type B8).
Probably this cluster is also part of the Sco-Cen OB association, since the space
velocities are similar.
 
\addtocounter{table}{4}
\begin{table*}[t]
\caption{\label{chab}Chamaeleon early-type stars}
\tiny{
\tabcolsep=0.08cm
\begin{tabular}{lcccrrrrrrrrrrrrrccrrc}
\hline \hline \\[-2.8pt]
 Name  & HIP & $l$ & $b$ & \multicolumn{1}{c}{d} & \multicolumn{1}{c}{X} & \multicolumn{1}{c}{Y} & \multicolumn{1}{c}{Z} &  $\mu_{l}{\rm cos}b$ & \multicolumn{1}{c}{$\mu_{b}$}  & $v_{\rm rad}$ & ref. & \multicolumn{1}{c}{U} & \multicolumn{1}{c}{V} & \multicolumn{1}{c}{W} &  \multicolumn{1}{c}{$V$}  & ($V$-$I_c$)o & Sp.T. & log$T_{\rm eff}$ & $A_V$ & ${\rm BC}_V$  & log$\frac{L}{L_{\odot}}$ \\
   HD  &         & \multicolumn{2}{c}{$\overline{\hbox{\hskip 25pt}{[\degr]}\hbox{\hskip 25pt}}$} & [pc]  & \multicolumn{3}{c}{$\overline{\hbox{\hskip 18pt}{\rm [pc]}\hbox{\hskip 18pt}}$} & \multicolumn{2}{c}{$\overline{\hbox{\hskip 7pt}{\rm [mas\,a^{-1}]}\hbox{\hskip 7pt}}$}& [km\,s$^{-1}$]&      & \multicolumn{3}{c}{$\overline{\hbox{\hskip 8pt}{\rm (LSR) [km\,s^{-1}]}\hbox{\hskip 8pt}}$} & & & & & & & \\[2.8pt]
(1) & (2) & (3) & (4) & (5) & (6) & (7) & (8) & (9) & (10) & (11) & (12) & (13) & (14) & (15) & (16) & \multicolumn{1}{c}{(17)} & (18) & (19) & (20) & (21) & (22) \\[2.8pt]
\hline\\
75416  &  42637  &  292.4024 &  $-$21.6511 &    97 &   34 &   $-$83 &  $-$36 &   $-$38.9 &   $-$10.4 &  14.0 & [2]  &   $-$2.2 &  $-$12.0 &   $-$2.4 &  5.46 &  $-$0.08 &  B9IV  &   4.05 &  0.00 & $-$0.52 &     1.91\\
75591  &  42830  &  290.7389 &  $-$20.5251 &   212 &   70 &  $-$186 &  $-$74 &   $-$14.4 &   $-$15.9 & $-$13.0 & [2]  &   $-$9.8 &   16.8 &   $-$3.2 &  8.09 &   0.05 & B9.5V  &   4.01 &  0.16 & $-$0.30 &     1.51\\
77981  &  44192  &  289.0698 &  $-$18.1660 &   196 &   61 &  $-$176 &  $-$61 &   $-$32.2 &   $-$14.2 &       &      &        &        &        &  6.85 &  $-$0.04 &   B9V  &   4.02 &  0.01 & $-$0.37 &     1.91\\
82423  &  46148  &  293.8218 &  $-$20.0690 &   192 &   73 &  $-$165 &  $-$66 &   $-$30.9 &    $-$3.4 &       &      &        &        &        &  7.85 &   0.07 & B9.5V  &   4.01 &   0.20 & $-$0.30 &     1.53\\
83979  &  46928  &  295.5743 &  $-$21.0438 &   165 &   67 &  $-$139 &  $-$59 &   $-$34.5 &   $-$14.9 &  10.0 & [3]  &  $-$12.1 &  $-$11.0 &   $-$7.3 &  5.07 &  $-$0.13 &   B5V  &   4.18 &  0.04 & $-$1.23 &     2.82\\
86320  &  48320  &  295.4593 &  $-$19.9209 &   216 &   87 &  $-$183 &  $-$74 &   $-$12.6 &     9.2 &  12.1 & [1]  &    4.6 &  $-$13.5 &   11.9 &  6.47 &   0.12 &  B8IV  &   4.09 &  0.42 & $-$0.70 &     2.44\\
93845  &  52633  &  297.7350 &  $-$18.9624 &   111 &   49 &   $-$93 &  $-$36 &   $-$35.5 &   $-$12.9 &  22.6 & [2]  &    2.4 &  $-$20.4 &   $-$6.6 &  4.45 &  $-$0.19 & B2.5IV &   4.32 &  0.05 & $-$2.04 &     3.05\\
96675  &  54257  &  296.6163 &  $-$14.5688 &   164 &   71 &  $-$142 &  $-$41 &   $-$21.7 &    $-$8.0 &       &      &        &        &        &  7.68 &   0.17 &   B6V  &   4.14 &  0.58 & $-$0.98 &     1.89\\
97300  &  54557  &  297.0329 &  $-$14.9195 &   188 &   83 &  $-$162 &  $-$48 &   $-$19.6 &    $-$9.5 &       &      &        &        &        &  8.97 &   0.45 &   B9V  &   4.02 &  0.96 & $-$0.37 &     1.40\\
98672  &  55308  &  297.0324 &  $-$13.3327 &   163 &   72 &  $-$141 &  $-$38 &   $-$22.0 &    $-$9.4 &  23.2 & [1]  &    4.4 &  $-$21.1 &   $-$5.2 &  6.26 &   0.01 &   A0V  &   3.99 &  0.05 & $-$0.22 &     1.94\\
102065 &  57192  &  300.0261 &  $-$17.9966 &   168 &   80 &  $-$138 &  $-$52 &   $-$28.0 &   $-$14.1 &       &      &        &        &        &  6.61 &   0.10 &  B9IV  &   4.05 &  0.33 & $-$0.52 &     2.06\\
102293 &  57374  &  299.1258 &  $-$14.4764 &   198 &   93 &  $-$168 &  $-$50 &   $-$30.5 &   $-$17.0 &       &      &        &        &        &  7.89 &   0.10 & B9.5IV &   4.04 &  0.31 & $-$0.45 &     1.65\\
104174 &  58484  &  300.2089 &  $-$15.6249 &   112 &   54 &   $-$93 &  $-$30 &   $-$38.5 &   $-$16.9 &  13.0 & [2]  &   $-$2.6 &  $-$13.8 &   $-$5.0 &  4.88 &  $-$0.02 &  B9Vn  &   4.02 &  0.05 & $-$0.37 &     2.22\\
104237 &  58520  &  300.2265 &  $-$15.5918 &   116 &   56 &   $-$97 &  $-$31 &   $-$37.1 &   $-$13.5 &       &      &        &        &        &       &        &        &        &        &        &         \\
106477 &  59758  &  300.2617 &  $-$10.3773 &   201 &  100 &  $-$171 &  $-$36 &   $-$20.0 &    $-$7.9 &       &      &        &        &        &  8.44 &   0.32 &   B9V  &   4.02 &  0.71 & $-$0.37 &     1.57\\
106911 &  60000  &  301.3369 &  $-$16.5431 &    83 &   41 &   $-$68 &  $-$24 &   $-$39.4 &     6.8 &  23.0 & [2]  &    8.6 &  $-$22.3 &    3.2 &  4.24 &  $-$0.11 &  B5Vn  &   4.18 &  0.08 & $-$1.23 &     2.57\\
115088 &  64951  &  304.1755 &  $-$17.1704 &   136 &   73 &  $-$108 &  $-$40 &   $-$20.5 &   $-$20.8 &       &      &        &        &        &  6.34 &  $-$0.01 &   A0V  &   3.99 &  0.02 & $-$0.22 &     1.73\\
116579 &  65628  &  305.3432 &  $-$11.9882 &   152 &   86 &  $-$121 &  $-$32 &   $-$37.1 &    $-$8.3 &  $-$3.0 & [2]  &  $-$14.2 &   $-$6.8 &    1.9 &  6.62 &  $-$0.03 &   B9V  &   4.02 &  0.03 & $-$0.37 &     1.79\\
123781 &  69576  &  307.8490 &  $-$14.3855 &   218 &  130 &  $-$167 &  $-$54 &   $-$15.9 &     1.5 &       &      &        &        &        &  7.83 &   0.02 &  B9IV  &   4.05 &  0.18 & $-$0.52 &     1.73\\
124654 &  70001  &  308.5824 &  $-$13.2816 &   192 &  117 &  $-$146 &  $-$44 &   $-$31.9 &    $-$4.1 &       &      &        &        &        &  7.72 &   0.07 &   B9V  &   4.02 &  0.23 & $-$0.37 &     1.63\\
124771 &  70248  &  306.9338 &  $-$17.9513 &   169 &   97 &  $-$129 &  $-$52 &   $-$15.3 &    $-$9.2 &   4.5 & [2]  &    1.4 &   $-$3.7 &   $-$1.2 &  5.06 &  $-$0.09 &   B4V  &   4.23 &  0.16 & $-$1.50 &     3.00\\[2.8pt]
\hline\\
\end{tabular}
{\vskip -5 pt}
References: [1] Andersen \& Nordstr\"om (\cite{andersen}); [2] Duflot et al. (\cite{duflot}); [3] van Hoof (\cite{vanhoof}).
}
\end{table*} 

\section{Spatial distribution} 
We present in Fig. 1 the distribution in galactic coordinates of both samples
(PMS stars and young early-type stars). The galactic coordinates of all stars
are listed in Cols. 2--3 of Table 1, Cols. 5--6 of Table 2, Cols. 3--4 of Tables
3, 4 and 5. In Table 1, Col. 4 also lists the references for the PMS stars coordinates.
All coordinates for the other stars are from the Hipparcos Catalogue.
It can be seen that the young stars, both PMS and young early-type, present a 
relatively continuous distribution. 
Some regions near molecular clouds, like $\rho$ Ophiuchi, Lupus or Chamaeleon, 
seem to have a larger fraction of low-mass stars, but nevertheless, some PMS stars 
can be found far from these regions that are usually considered as the main SFRs.     

  \begin{figure}
   \resizebox{\hsize}{!}{\includegraphics{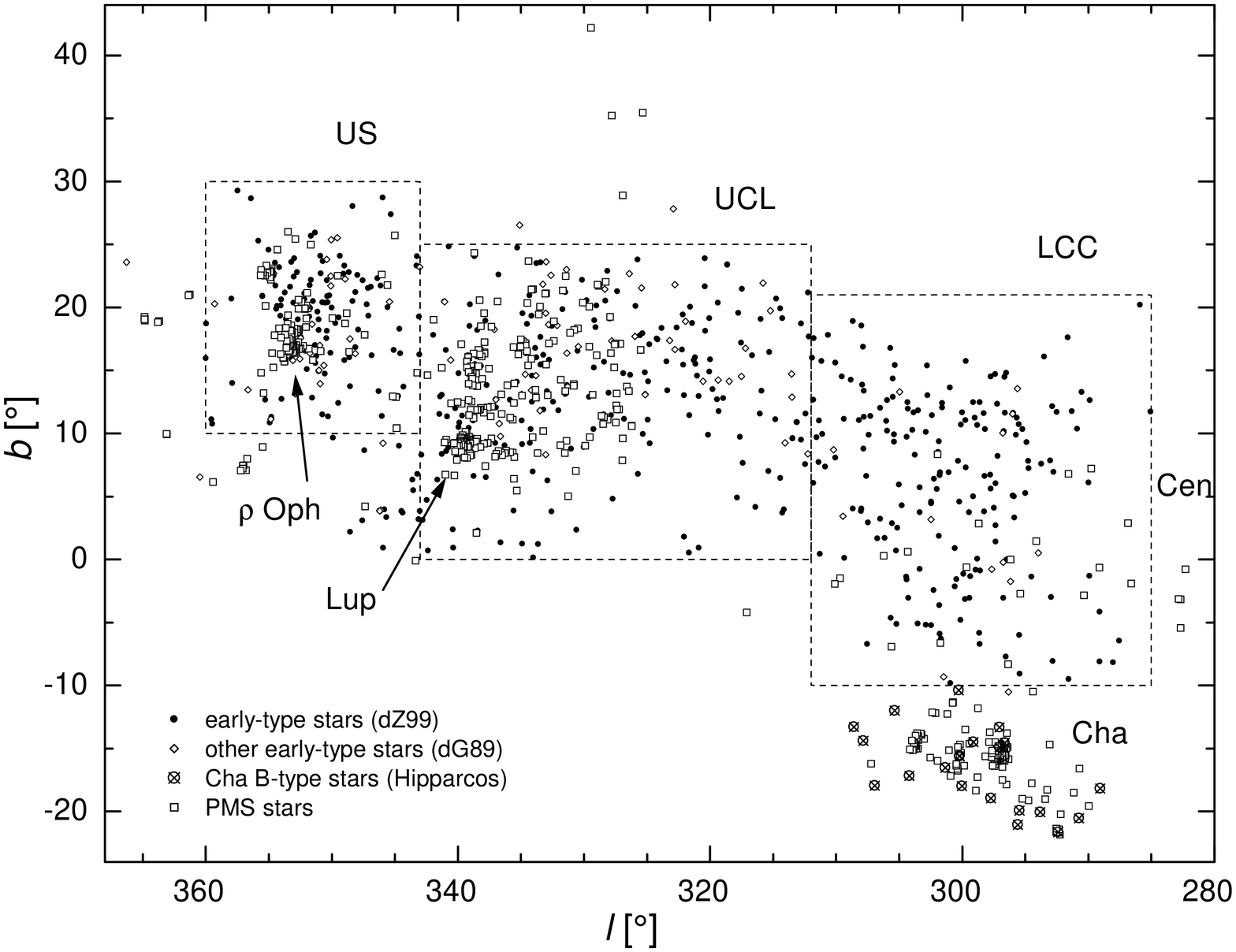}}
      \caption{Distribution in galactic coordinates of the PMS stars and  
	  of the stars of the OB associations, listed by dZ99 and other  
	  massive stars selected in this work (probable Sco-Cen members and 
	  Chamaeleon B-type stars).
	  }
   \end{figure}

The range of distance of the objects of interest extends to about 250 pc, but 
most of the young stars and molecular clouds are situated at distances between
100 and 200 pc. 
If we compare the mean distances of the molecular clouds (see values in Sect. 5)
with the mean distances of the subgroups of the Sco-Cen OB association (dZ99), 
we may think that the PMS stars do not occupy the same region of space of the 
young early-type stars. However, when we analyze the individual distances of 
these young stars we verify that that they are spread over a large range of 
distances. Sartori et al. (\cite{sart01}) noted that their distribution extends to 
distances as small as 60 pc, because there are at least 10 PMS stars and several 
young early-type stars that have distances less than 100 pc.
The individual distance values and the errors derived from the parallax 
errors are listed in Cols. 9--11 of Table 2 and Cols. 5--7 of Tables 3 and 4. 
In Table 2 we also give the parallax values and respective errors (Cols. 7--8) 
in order to show that Hipparcos did not obtain reliable parallaxes to some PMS 
stars (9 stars in the studied SFRS) and therefore, their distances could not be 
estimated. 

Then, we calculated the positions in the Galactic system (X, Y, Z) based on these 
distances derived from the Hipparcos parallaxes. The positions in the Galactic 
system are listed in Cols. 12--14 of Table 2, Cols. 8--10 of Tables 3 and 4, 
and Cols. 6--8 of Table 5. 
The distribution in 3 dimensions of these positions is shown in Fig. 2. In this 
diagram, as only the stars that have reliable Hipparcos parallax are represented, 
not more than 30 PMS stars are shown. However, it can be clearly seen that the young 
early-type stars and the PMS stars follow a similar spatial distribution. There
is no separation between the PMS and the young early-type stars.      

It is not usual in the literature to consider that the Chamaeleon SFR is related 
in some way to the Ophiuchus and Lupus SFRs. We consider that they are part of 
the same complex because there are in the region between Ophiuchus and 
Chamaeleon other young stars, like the PMS stars in Centaurus, and also molecular 
clouds, like the Musca clouds, the Coalsack, etc. (all at about the same distance).

  \begin{figure}
   \resizebox{\hsize}{!}{\includegraphics{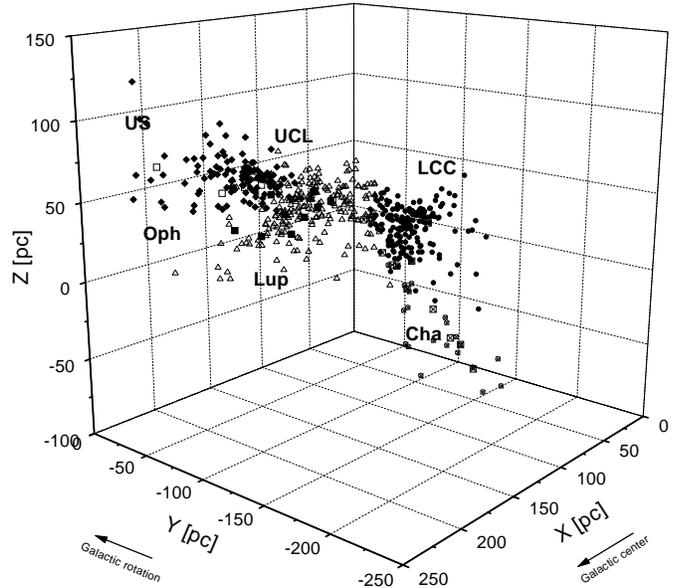}}
      \caption{~3D-distribution of positions in the Galactic system of the stars 
	  with Hipparcos parallaxes: Ophiuchus (open squares), Lupus (filled squares), 
	  and Chamaeleon (squares with times) PMS stars, and US (filled diamonds), 
	  UCL (open triangles), LCC (filled circles), and Cha (circles with times) 
	  early-type stars. The X axis points towards the Galactic center and the 
	  Y axis towards the direction of Galactic rotation.
	  }
   \end{figure}  

We show in Fig. 3 the projected positions of the young stars on the XY plane. 
Interestingly, in the XY distribution of stars we can distinguish three denser groups, 
separated by gaps of lower stellar density. These groups are the US, UCL and 
LCC associations. The elongated shapes of these groups in the direction opposite 
to the Sun is largely an effect of the errors of measurements on distances. In Fig. 4 
we present the error bars in position in the X and Y directions, derived from the 
Hipparcos errors on parallaxes. Probably most of the stars that seem to be at 
distances larger than 200 pc are, in reality, closer than that, considering that 
the errors increase quickly with the distance to the Sun. The errors are expected 
to produce an excess density of star in the direction that connects the Sun to the 
main concentrations of stars, as observed. The separation between the three groups 
seems to be real, and not an effect of extinction due to molecular clouds, since 
the denser parts of the known clouds do not coincide with the gaps. The apparent 
geometry of the distribution of stars in the XZ plane (Fig. 5), in which the distance 
of the stars to the Galactic plane increases with their distance to the Sun, is 
probably due to the combined effect of the errors in distance (that tends to produce 
elongated structures) and of a larger extinction in the Galactic plane. 
The well-known fact that the Sun is about 20 pc above the Galactic plane can 
also be noticed in the diagram of Fig. 5.

   \begin{figure}
   \resizebox{\hsize}{!}{\includegraphics{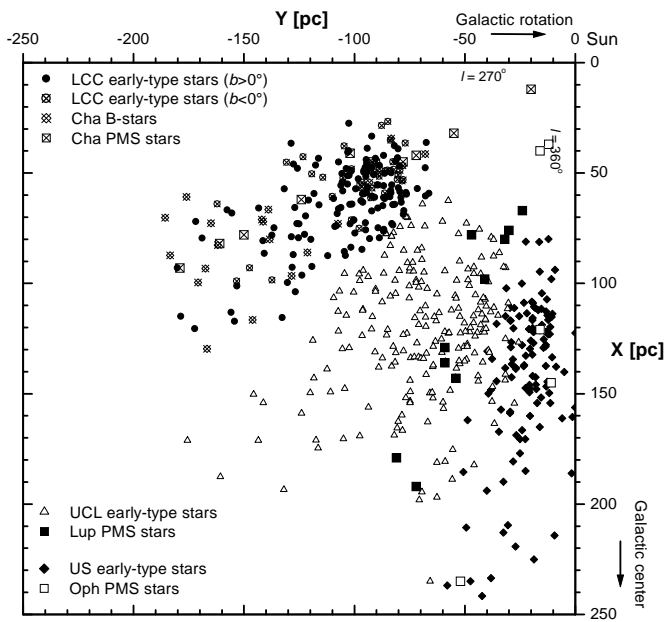}}
      \caption{Projection on the Galactic plane of the positions of PMS 
	  and young early-type stars. The X axis points towards the Galactic 
	  center and the Y axis towards the direction of Galactic rotation.}
   \end{figure}

   \begin{figure}
   \resizebox{\hsize}{!}{\includegraphics{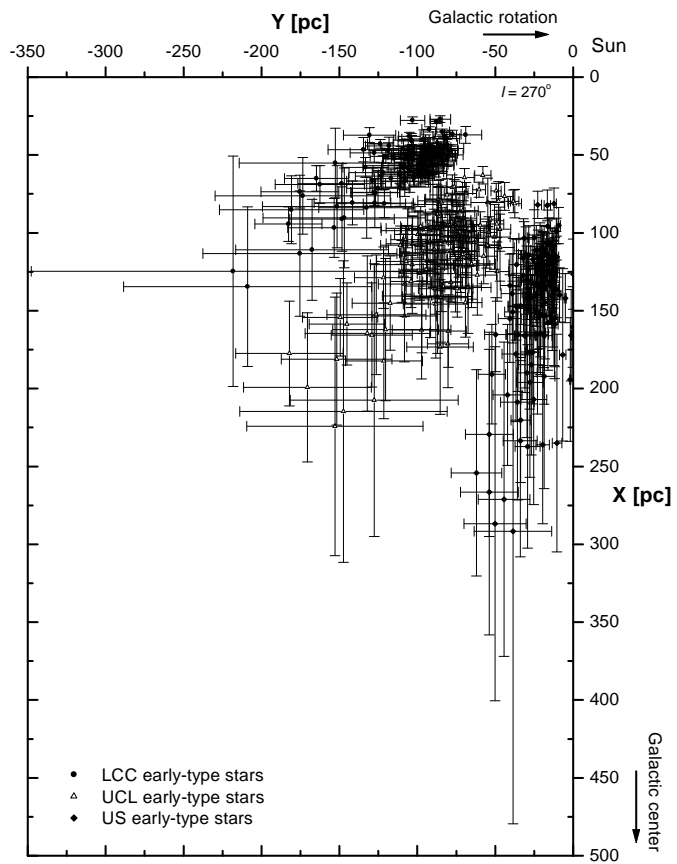}}
      \caption{Projection on the Galactic plane of the positions of the 
	  young early-type stars. The error bars in the X and Y positions are
	  due to the Hipparcos parallax errors. The axis orientation is the
	  same of Fig. 3.}
   \end{figure}
    
   \begin{figure}
   \resizebox{\hsize}{!}{\includegraphics{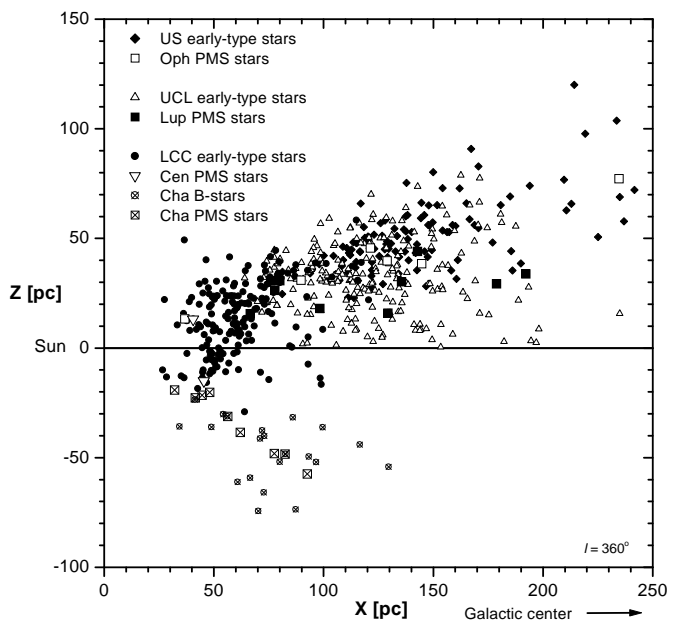}}
      \caption{Projection of the positions of PMS and young early-type stars on 
	  the XZ plane. The X axis points towards the Galactic center.}
   \end{figure}      

\section{Kinematics}

\subsection{Proper motions}

We collected in the literature all the available proper motions of the young stars 
in the studied region. The Hipparcos Catalogue (ESA \cite{esa}) provides accurate 
measurements of proper motions for all the 521 young early-type stars of the Sco-Cen 
OB association (dZ99) and also for the 88 early-type stars from dG89 that we 
selected to analyze in this work. We adopted the Hipparcos proper motions for the 
39 PMS that also are in the catalogue. For the other PMS stars, we mainly adopted 
the proper motions determined by Teixeira et al. (\cite{teixeira}; data for 160 
stars). The work by Teixeira et al. and some others (Preibisch et al. \cite{preib98}; 
Frink et al. \cite{frink}; Terranegra et al. \cite{terranegra}) were done before 
the Tycho-2 Catalogue (Hog et al. \cite{hog}) became available.
However, as the determination of proper motions done by these works were specifically 
for PMS stars, there are proper motions for several stars exclusively determined by
them. Our list of stars with known proper motions was completed with 9 stars from 
the Tycho-2 Catalogue and 6 from Terranegra et al. (\cite{terranegra}). 
The proper motions in galactic coordinates are listed in Cols. 5--6 of Table 1 
(references are in Col. 7), in Cols. 15--16 of Table 2, in Cols. 11--12 of Tables 3 
and 4, and in Cols. 9--10 of Table 5. 
We present in Fig. 6 the vector proper motion diagrams (VPDs) in galactic
coordinates of the early-type stars of the three OB associations, US, UCL and LCC, 
and of the Chamaeleon association (represented by circles), and of the PMS stars of 
Ophiuchus, Lupus and Chamaeleon SFRs.  

   \begin{figure*}
  \includegraphics[width=18cm]{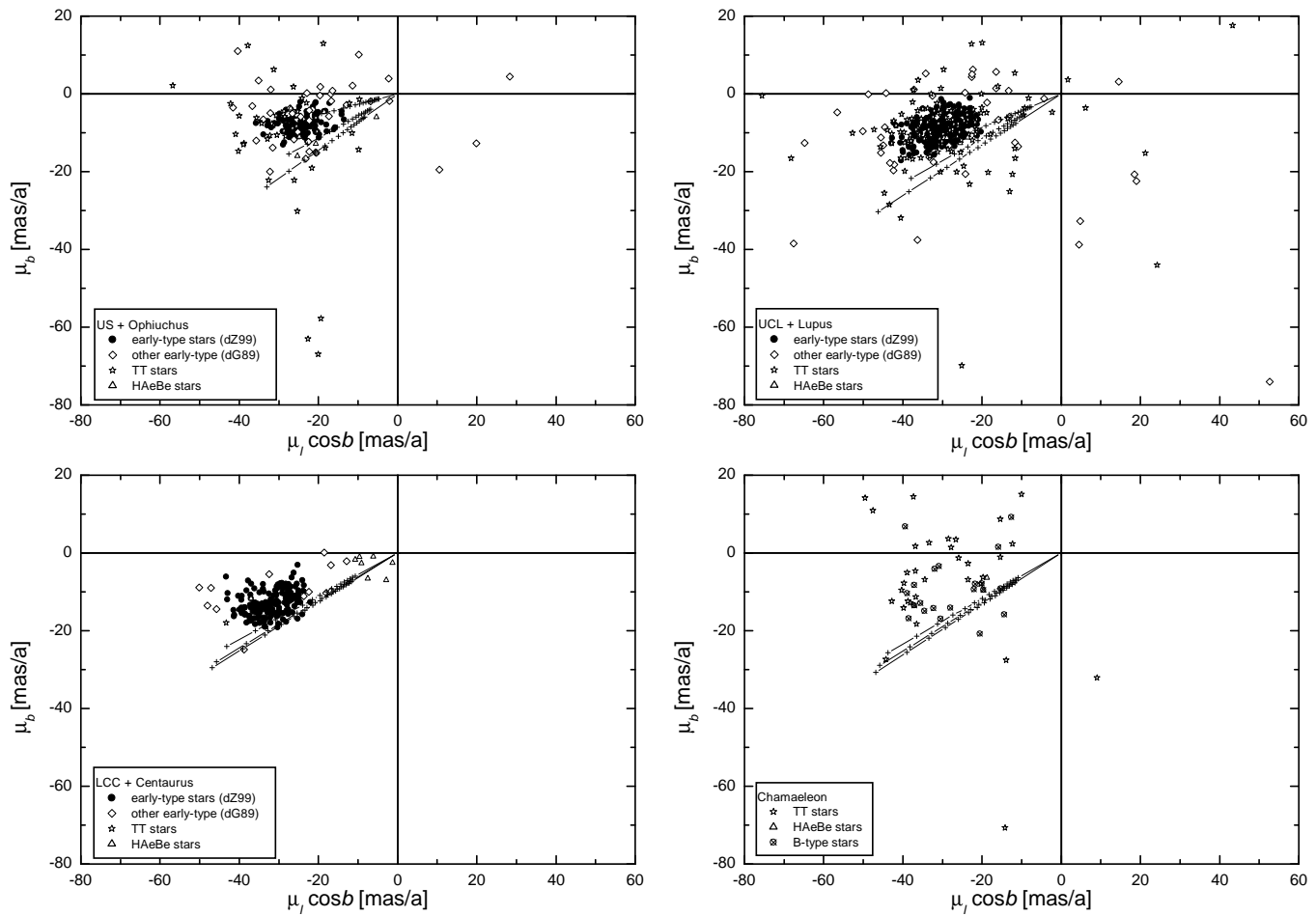}
      \caption{VPD in galactic coordinates of the PMS stars and of the early-type 
	  stars selected by dZ99, the other early-type stars from dG89 excluded by dZ99,
	  and the Chamaeleon B-type stars, from: {\bf a)} US and Ophichus; 
	  {\bf b)} UCL + Lupus; {\bf c)} LCC + Centaurus; {\bf d)} Chamaeleon.
	  The small crosses connected by solid lines indicate the values of the reflex
	  of Solar motion as a function of distance, from 50 to 200 pc (in steps of 10 pc),
	  for two extreme directions and for the mean direction of each field.
	  }
   \end{figure*}

When comparing the proper motions of the young early-type and the PMS stars samples, 
it is important to discuss an interesting selection effect in the OB sample. The 
proper motions of the early-type stars selected by dZ99 are much more ordered (all 
similar) than those of the PMS stars. However, this is largely due to the fact that 
dZ99 only considered stars that were in a very narrow range of proper motion 
components, on both axes, as members of the Sco-Cen OB association. On the other
hand, we have analyzed all the PMS stars with known proper motions, i.e., they were 
not selected by their kinematic properties. 
In Fig. 6, we show separately the proper motion of the dG89 stars that were 
selected for comparison. Obviously, these stars present a larger scatter of their 
proper motions than those selected by dZ99. However, they are at the same average 
distance and have the same ages (as later discussed in Sect. 5) of the stars 
considered as members by dZ99. 
If we consider the total sample of young early-type stars (the stars selected by 
dZ99 plus the re-inserted ones), then the scattering of proper motions of the 
young early-type stars are similar to that of the PMS stars. The VPD of the 
Chamaeleon region (Fig. 6.d) shows that the B-type and the PMS stars have also very 
similar proper motion distributions. The whole sample of young stars exhibit a 
common direction of motion, as can be seen in Fig. 7. 

We note that some stars selected by dZ99 were previously classified as PMS stars 
in other studies: HD 145718 and HD 81624 in US (Gregorio-Hetem et al. \cite{gregorio}; 
Th\'e et al. \cite{the}), HD 138009, HD 141277, and HD 143677 in UCL (Brandner et al. 
\cite{brandner}; Krautter et al. \cite{krautter}), and HD 100546 in LCC (Torres 
\cite{torr99}). This means that the kinematics of at least a few PMS stars is exactly 
the same of the Sco-Cen OB associations members selected by dZ99. Considering that 
only a small number of PMS stars belongs to the source of data used by dZ99, the 
Hipparcos Catalogue, we believe that it is not a coincidence, i.e., those PMS stars 
are not interlopers. On the contrary, it reinforces the idea that the PMS and the 
early-type stars show a very similar kinematics. 

   \begin{figure}
   \resizebox{\hsize}{!}{\includegraphics{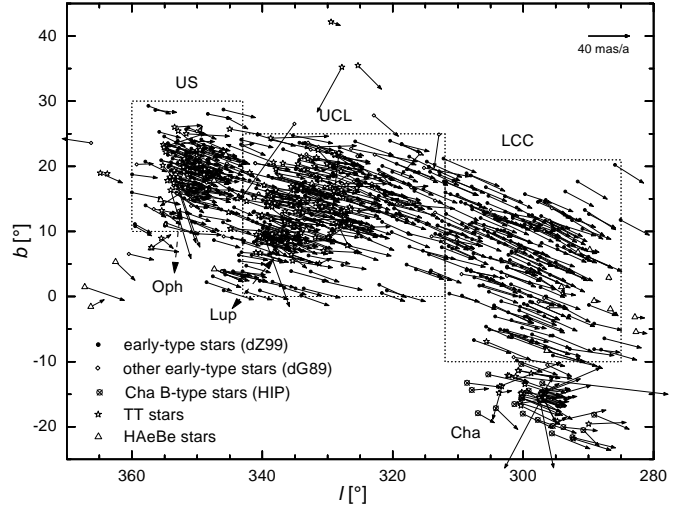}}
      \caption{Proper motions vectors, in galactic coordinates, of all the young 
	  early-type and PMS stars of the studied regions.
	  }
   \end{figure}

A valid question is whether the common proper motion can be simply the 
reflex motion due to Solar velocity with respect to the local standards of rest, 
as the stars are in a relatively narrow range of distances. 
Since the effect of Solar motion on the star proper motion depends on its 
distance, and we know the distance of few PMS stars, we show in the diagrams 
of Fig. 6 the values of the reflex Solar motion as a function of distance, from 
50 to 200 pc. We assumed the basic Solar motion, with components: U$_\odot$ = 
10.0 km\,s$^{-1}$ (radially towards the Galactic center), V$_\odot$ = 
5.25 km\,s$^{-1}$ and W$_\odot$ = 7.17 km\,s$^{-1}$ (Dehnen \& Binney \cite{dehnen}).
The reflex of Solar motion depends also on the direction of the stars, and since the 
regions studied here have sizes of several degrees, we present it for two extreme 
directions and for the mean direction of each field. It can be seen that the
reflex Solar motion for distances less than 150 pc represents a significant 
fraction of the observed proper motions of these stars. 
The corrected proper motion vectors of the young stars with known distances are 
shown in Fig. 8. Note that, although the values of the corrected proper motions 
are much smaller than the observed ones, a common direction of motion persists.  

   \begin{figure}
   \resizebox{\hsize}{!}{\includegraphics{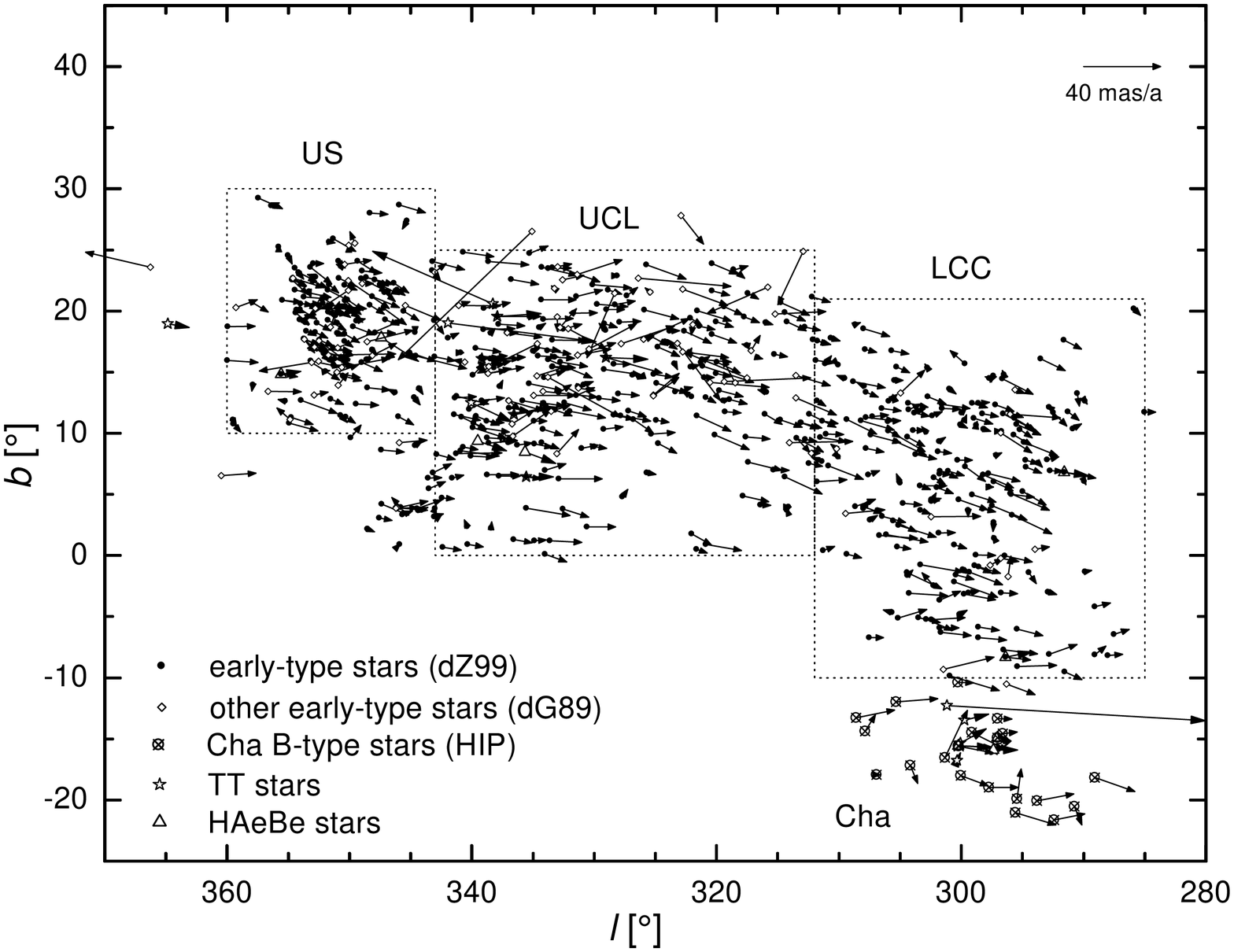}}
      \caption{Proper motions vectors, in galactic coordinates, corrected for 
	  Solar motion, of the young early-type and the PMS stars with known distances.
	  } 
   \end{figure}
   
\subsection{Radial velocities}

We performed a systematic search in the literature for radial velocities of the 
Hipparcos stars. The $v_{\rm rad}$ values and their references are listed in Cols. 
17--18 of Table 2, Cols. 13--14 of Tables 3 and 4, and Cols. 11--12 of Table 5. 
In Fig. 9 we show the histograms of the $v_{\rm rad}$ of the early-type stars of 
each studied OB associations in comparison with the $v_{\rm rad}$ of the PMS stars
of the related SFRs. It can be remarked that the average $v_{\rm rad}$ vary with 
longitude, as expected from the projection of the Solar motion, as discussed in 
more detail in next sub-section. 
Although the sample of PMS stars with measured $v_{\rm rad}$ is small, no difference 
can be noticed between the velocities of PMS and young early-type stars. 

   \begin{figure}
   \resizebox{\hsize}{!}{\includegraphics{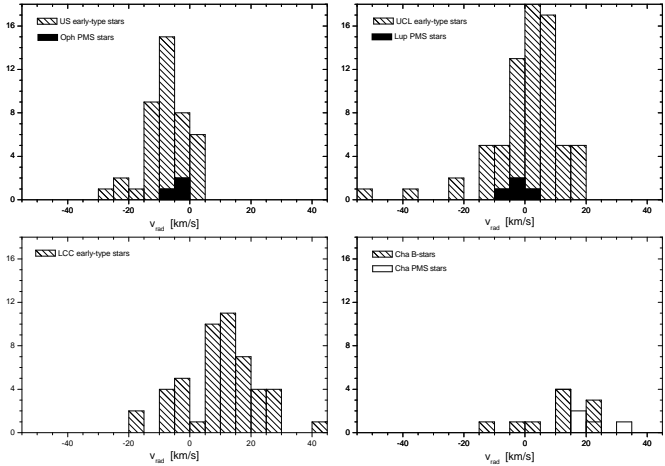}}
      \caption{Histograms of radial velocities of the PMS and the early-type stars
	  from: {\bf a)} US and Ophichus; {\bf b)} UCL and Lupus; {\bf c)} LCC; 
	  {\bf d)} Chamaeleon.
	  }
   \end{figure}

\subsection{Space velocities}

We combined the proper motions and radial velocities of the stars with the
distances derived from Hipparcos parallaxes, to calculate the U, V, W components 
of the space velocities in the Galactic frame. We corrected the space velocities   
for the velocity of the Sun with respect to the local standard of rest (LSR). 
Note that the effect of the LSR correction is to produce a shift in the space 
velocities; it does not affect the aspect of the velocity distributions.
The U, V and W components with respect to the LSR are listed in Cols. 19--21 
of Table 2 for the Hipparcos PMS stars and in Cols. 15--17 of Tables 3 and 4,
and in Cols. 13--15 of Table 5, for the early-type stars.

In Table 6 we present the mean values and the standard deviations ($\sigma$), 
resulting from a Gaussian model, for each OB association. It can be seen that the 
velocity differences between the 4 associations are not significant, being smaller 
than 1 $\sigma$ of the individual distributions (of about 5 km\,s$^{-1}$).  
For each SFR the space velocities of the Hipparcos PMS stars are compatible with 
the mean values of the OB associations.

\begin{table}[ht]
\caption{\label{spvel} Space velocities of the studied associations: mean value 
and standard deviation ($\sigma$) resulting from a Gaussian model, in units of 
km\,s$^{-1}$.}
\scriptsize{
\tabcolsep=0.065cm
\begin{tabular}{lrcrcrcrcrcrc}
\hline \hline \\[-3pt]
Association & \multicolumn{2}{c}{U} & \multicolumn{2}{c}{V} & \multicolumn{2}{c}{W} & \multicolumn{2}{c}{U$_{\rm LSR}$} & \multicolumn{2}{c}{V$_{\rm LSR}$} & \multicolumn{2}{c}{W$_{\rm LSR}$} \\
             & {\tiny mean} & {\tiny $\sigma$} & {\tiny mean} & {\tiny $\sigma$} & {\tiny mean} & {\tiny $\sigma$} & {\tiny mean} & {\tiny $\sigma$} & {\tiny mean} & {\tiny $\sigma$} & {\tiny mean} & {\tiny $\sigma$} \\[3pt]
\hline \\[-3pt]
US  & -6.7 & {\tiny 5.9} & -16.0 & {\tiny 3.5} & -8.0 & {\tiny 2.7} & 2.4 & {\tiny 5.0} & -10.0 & {\tiny 3.3} & -0.9 & {\tiny 2.8} \\
UCL & -6.8 & {\tiny 4.6} & -19.3 & {\tiny 4.7} & -5.7 & {\tiny 2.5} & 3.2 & {\tiny 4.2} & -14.4 & {\tiny 4.6} & 1.4  & {\tiny 2.4} \\
LCC & -8.2 & {\tiny 5.1} & -18.6 & {\tiny 7.3} & -6.4 & {\tiny 2.6} & 1.1 & {\tiny 5.8} & -13.6 & {\tiny 5.6} & 1.1  & {\tiny 2.4} \\
Cha$^{\mathrm{a}}$ & -11.9 & {\tiny 7.7} & -16.0 & {\tiny 11.4} & -8.6 & {\tiny 5.8} & -1.9 & {\tiny 7.7} & -10.8 & {\tiny 11.4} & -1.4 & {\tiny 5.8} \\[3pt]
\hline
\end{tabular}
{\vskip 2.0pt}
$^{\mathrm{a}}$ For the Cha OB association the mean velocities and the $\sigma$ 
presented are simple statistics, not derived from Gaussian fits.
}
\end{table}

The average V component in the LSR is about -12 km\,s$^{-1}$, which is significant 
and contrary to the velocity expected from Galactic differential rotation. 
If the rotation curve is relatively flat, stars situated at a Galactic radius 
smaller than the Solar radius R$_{\odot}$ must complete a rotation in a shorter 
time than the period associated with the LSR, and should appear to us with a 
positive velocity in the direction of rotation.

Figure 10 shows the resultant velocities in the LSR of individual stars in the 
XY plane. As expected from the combination of the V component already discussed 
and of the small U component on the average, the young stars are moving away from 
the Sun.

   \begin{figure}
   \resizebox{\hsize}{!}{\includegraphics{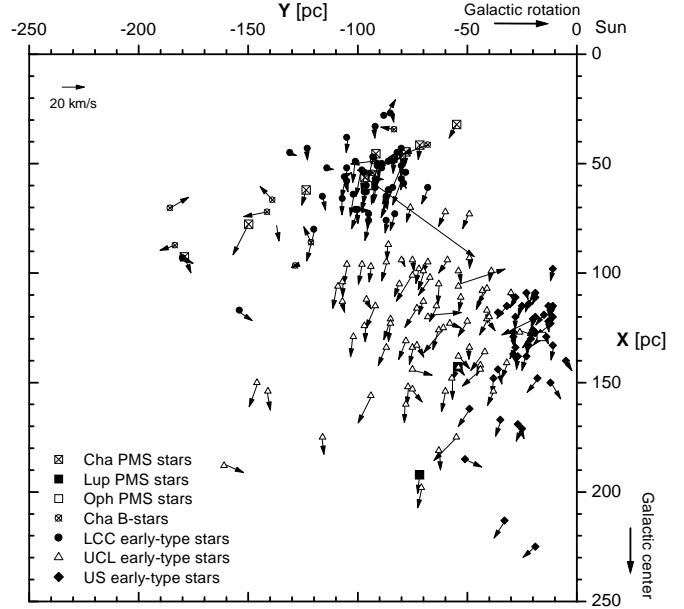}}
      \caption{U--V velocities resultant vectors in the XY plane of young 
	  early-type and PMS stars. The axis orientation are the same of Fig. 3. 
	  }
   \end{figure}

\section{H-R diagrams and age determination}

We estimated the values of effective temperature for the early-type and 
PMS stars from the spectral types and luminosity classes given in the 
literature, using the relationship between spectral type and effective 
temperature given by de Jager \& Nieuwenhuijzen (\cite{dejager}), who 
present a complete grid of spectral types and luminosity classes. 
In the cases where no luminosity class was given, we adopted the 
luminosity class V. The spectral types and luminosity classes and their 
references are listed in Cols. 11--12 of Table 1, 25--26 of Table 2 
and 20 of Tables 3 and 4 (for the early-type stars, the spectral types 
were adopted exclusively from the Hipparcos Catalogue). The log$T_{\rm eff}$ 
values are listed in Cols. 13 of Table 1, 27 of Table 2, and 21 of 
Tables 3 and 4. 

We calculated the visual extinction $A_V$ towards the individual stars 
using their observed ($V$-$I_c$) color index and the intrinsic color index 
from Bessel et al. (\cite{bessel}; their Tables 1, 2 and 3) and the relations 
of $A_V$ and E($V$-$I_c$) of Schultz \& Wiemer (\cite{schultz}). Bessel et al. 
(\cite{bessel}) give intrinsic colors as a function of $T_{\rm eff}$ and log\,$g$. 
In order to get the intrinsic colors for each star, we adopted the log\,$g$ 
derived from its spectral type and luminosity class as given by Table VII of 
Strai\v{z}ys \& Kuriliene (\cite{straizys}) (the adopted log\,$g$ values are 
listed in Cols. 14 of Table 1, 28 of Table 2, and 22 of Tables 3 and 4). 
The observed and intrinsic ($V$-$I_c$) color index and the $A_V$ values are 
listed in Cols. 9, 15 and 16 of Table 1, 23 and 29 of Table 2 (we do not give 
the intrinsic colors for the Hipparcos PMS stars), and 19, 23 and 24 of Tables 3 
and 4. The ($V$-$I_c$) references for the PMS stars are listed in Col. 10 of 
Table 1 and for the Hipparcos PMS stars, in Col. 24 of Table 2. For the early-type 
stars, the observed ($V$-$I_c$) were also extracted exclusively from the Hipparcos 
Catalogue. 

Then, bolometric luminosities were calculated through the $V$
magnitude corrected for interstellar extinction, the visual bolometric 
correction BC$_V$, also given by Bessel et al. (\cite{bessel}), and the distances
of each star, when available, or, otherwise, the mean distance of the SFR.
We adopted the measured parallaxes to derive the distances of the PMS and 
early-type stars that are in the Hipparcos Catalogue. The SFR adopted mean 
distances, also derived from Hipparcos, were 168 pc for Cha I, 147 pc for 
Lupus, 128 pc for $\rho$ Ophiuchi (Bertout et al. \cite{bertout}), 178 pc for 
Cha II (Whittet et al. \cite{whittet}) and 140 pc for US (dZ99). The $V$ 
magnitudes, the BC$_V$ and the log($L$/$L_{\odot}$) values are listed in Cols. 8, 
19 and 20 of Table 1 for 240 PMS stars, in Cols. 22 and 30 of Table 2 for 28
Hipparcos PMS stars (we do not give the BC$_V$ values for the Hipparcos PMS 
stars), and in Cols. 18, 25 and 26 of Tables 3 (485 early-type stars) and 
4 (87 other early-type stars). The references for the $V$ magnitudes are the 
same for the ($V$-$I_c$) color indexes. 

The H-R diagrams are presented in Figs. 11 to 14, separately for each of the 
main SFRs. The error bars in luminosity correspond to the errors in the
Hipparcos parallaxes. 
The theoretical isochrones of Siess et al. (\cite{siess}) for the PMS phase, and 
of Bertelli et al. (\cite{bertelli}) for the evolution after the main sequence, 
are shown in these figures. We adopt the isochrones of Siess et al. (\cite{siess}) 
because they are computed for stars in the mass range 0.1 to 7.0 $M_{\odot}$, which 
is essential to represent not only the PMS phase of the TT stars but also of the 
HAeBe stars. 
Sartori (\cite{sart00}) compared this set of evolutionary tracks with others 
usually adopted in the literature (Palla \& Stahler \cite{palla}; D'Antona \& 
Mazzitelli \cite{dantona}; Baraffe et al. \cite{baraffe}), and concluded that the 
grids of PMS tracks of Siess et al. (\cite{siess}) are adequate for the purposes 
of this work.
The ages of the PMS isochrones shown in Figs. 11 to 14 are between 0.1 and 20 Myr 
(the numbers at the bottom of the curves correspond to the ages, in units of Myr). 
The zero age main sequence (ZAMS) curve computed by Siess et al. (\cite{siess}) 
is also shown. 
The isochrones of Bertelli et al. (\cite{bertelli}) are derived from stellar 
models computed from the ZAMS to the central carbon ignition for massive stars 
or to the beginning of the thermally pulsing regime of the asymptotic giant branch 
phase for low and intermediate mass stars. In these figures, the ages of the 
isochrones of Bertelli et al. (\cite{bertelli}) range from 8 to 25 Myr (the numbers 
at the end of the curves correspond to the ages, in units of Myr).

Both sets of theoretical isochrones adopted are computed for the Solar composition. 
Although there are very few measurements of the PMS stars abundances (e.g. Padgett 
\cite{padgett}) that could confirm this assumption, the Solar composition is usually 
adopted for all close SFRs. Sartori (\cite{sart00}) investigated the abundance 
measurements for the Sco-Cen OB association members, and concluded that there is 
not enough data to assert that the metallicity of these young stars is different 
from the Solar metallicity. 

   \begin{figure}
   \resizebox{\hsize}{!}{\includegraphics{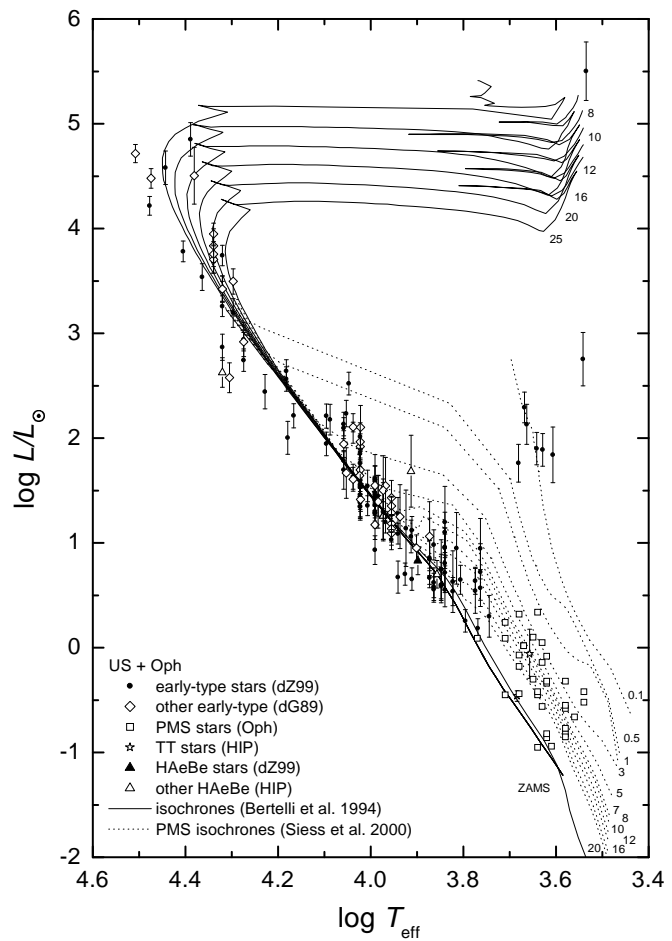}}
      \caption{H-R diagram of the stars of the US OB association and of the PMS 
	  stars of Ophiuchus. The error bars in luminosity correspond to the errors 
	  in parallax. The isochrones ages, in units of Myr, are indicated 
	  by the numbers at the bottom of the PMS isochrones curves, and 
	  at the end of the curves, for the evolution after the main sequence.
	  }
   \end{figure}

   \begin{figure}
   \resizebox{\hsize}{!}{\includegraphics{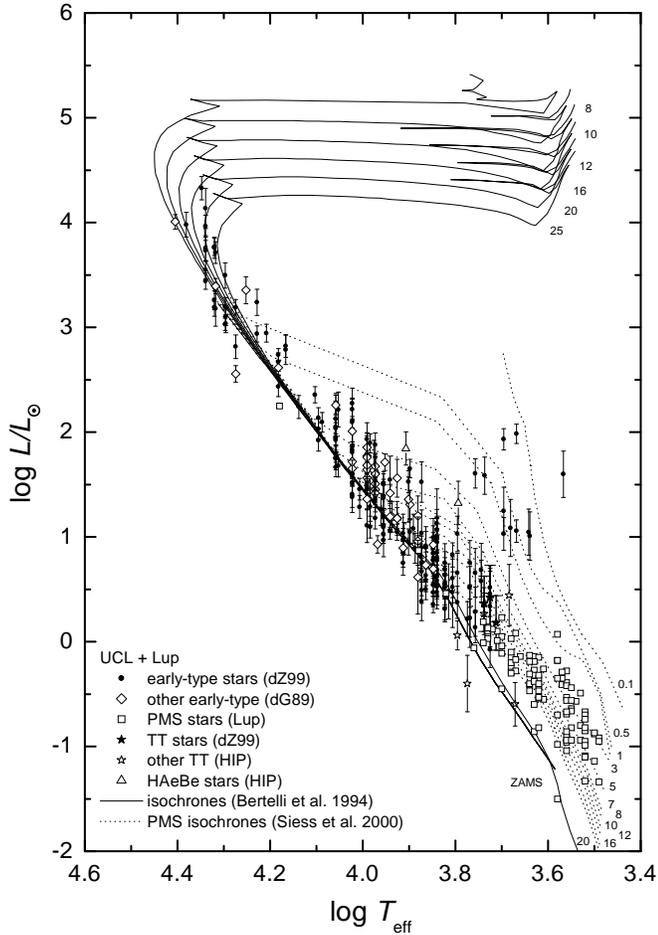}}
      \caption{H-R diagram of the stars of the UCL OB association and of
	  the PMS stars of Lupus (see details in the caption of Fig. 11).   
	  }
   \end{figure}

   \begin{figure}
   \resizebox{\hsize}{!}{\includegraphics{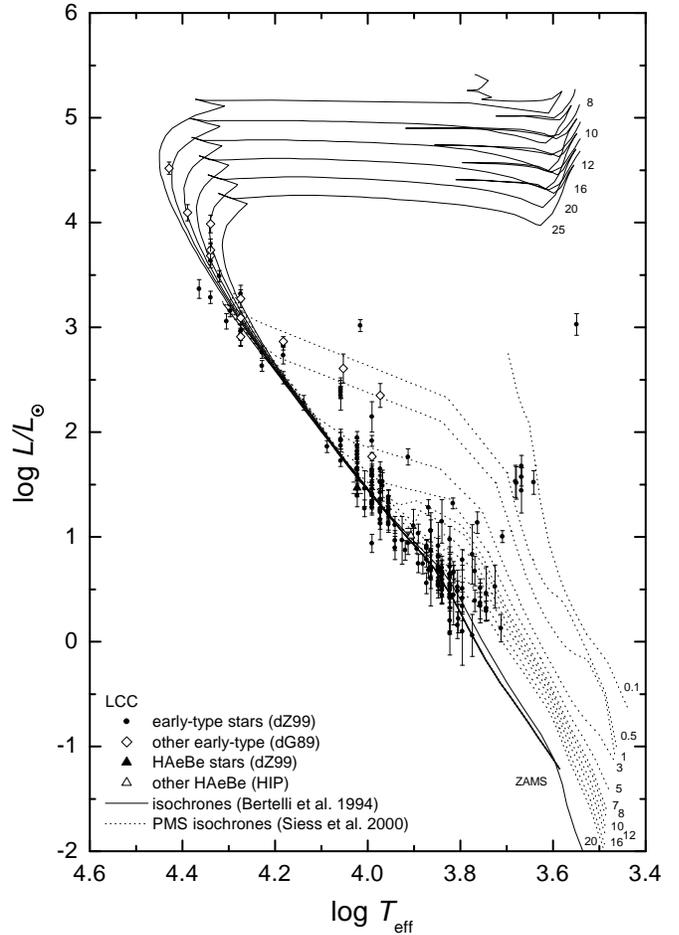}}
      \caption{H-R diagram of the stars of the LCC OB association
	   (see details in the caption of Fig. 11).
	   }
   \end{figure}

   \begin{figure}
   \resizebox{\hsize}{!}{\includegraphics{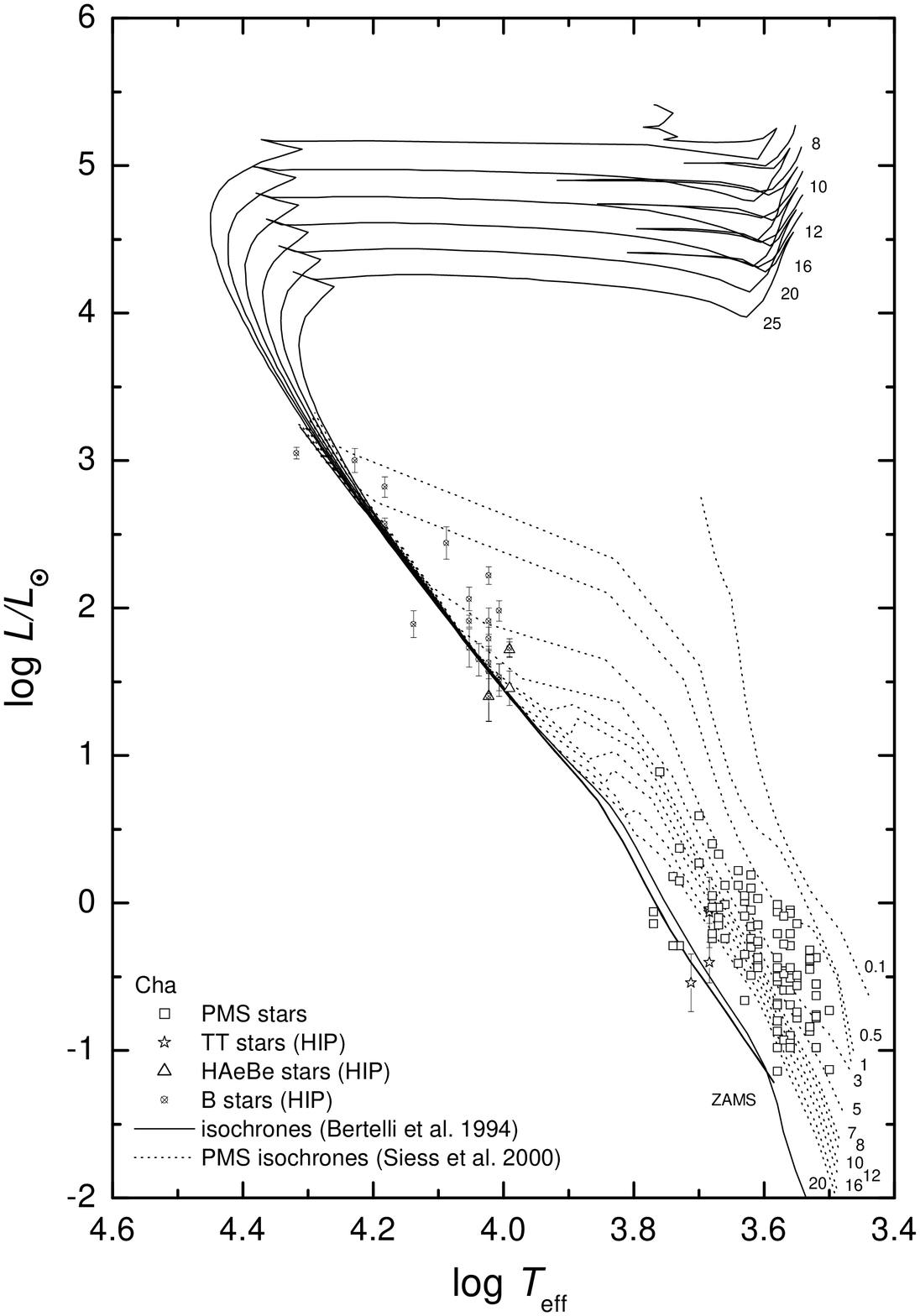}}
      \caption{H-R diagram of the PMS stars of Chamaeleon and of the
	  Hipparcos B-type stars (see details in the caption of Fig. 11). 
	  }
   \end{figure}   

The main conclusions that we draw from the H-R diagrams are the following:
 \begin{enumerate}
\item  In each SFR the PMS stars span a wide range of ages, from 1 to 20 Myr. 
Although Preibisch \& Zinnecker (\cite{preib99}) and Preibisch et al. (\cite{preib02}), 
analyzing the PMS stars in the US region, concluded that they have an age of 5 Myr, 
without a significant spread, we can observe in their H-R diagram many objects 
above the 1 Myr isochrone, and many objects close to the 10 Myr isochrone. 
It seems that their conclusions are different from ours because they are based 
on a different interpretation of the H-R diagrams.

\item For the OB associations, the ages that we obtain from a comparison 
of the positions of the early-type stars on the diagrams with the isochrones
are 8--10 Myr for US, 16--20 Myr for UCL and for LCC. For the Chamaeleon B
association, due to the absence of sufficiently massive stars, it is
not possible to determine its age. 
These ages are different from those determined by dG89 (5--6 Myr for US,
14--15 Myr for UCL, 11--12 Myr for LCC). The discrepancies are explained by 
the fact that we use Hipparcos distances, more recent isochrones, and 
our effective temperatures are derived from the spectral type (in the work 
of dG89 the luminosity determination is based exclusively on photometric 
measurements). It is important to emphasize that in our method as well as 
in that of dG89, the age determination depends on a few stars situated in 
the upper part of the H-R diagram, so that the uncertainties are large. 
Our results do not confirm that UCL is older than the LCC 
association, as found by dG89. 

\item The stars considered as members of the associations by dG89, but
excluded by dZ99, as discussed in Sect. 2.2, are indeed members, if we 
consider their position on the H-R diagrams (they are all in the main sequence 
-- see Figs. 11 to 13). 

\item On the other hand, the G- to M-type stars selected as members are 
actually interlopers, since they are old stars already in the red giant branch. 
One exception is $\alpha$ Sco, situated in the upper part of the H-R diagram 
of US, which is a member.
Although in principle stars of spectral types G to M and classified as luminosity 
class III could be PMS stars, these stars of the dZ99 samples are not PMS because
they are too bright to be unknown TT stars. 
 \end{enumerate}

Although we estimated ages in the range 1--20 Myr for the PMS stars of US, 
and ages in the range of 8--10 Myr for the US OB association, we cannot say 
that there is indeed a difference in the age of the two populations.
Note that Preibisch et al. (\cite{preib02}) estimated an age of 5 Myr for the 
early-type stars of US, similar to the age of the PMS stars.
In our analyses, we could not see any clear evidence that in the process
of star formation, massive stars form first.

Finally, in all the diagrams we observe a number of stars situated on the 
left of or below the ZAMS\@. Several types of errors could contribute to this; 
for instance the distance or spectral type could be incorrect, some stars 
could be binaries, others could be interlopers with low metallicity. 

\section{Discussion}

The spatial distribution analysis of the studied young stellar associations has
shown that they form an almost continuous structure. Sartori et al. (\cite{sart01}) 
noted that possibly other close groups of young stars could also be part of this 
large structure.

There is a clear kinship between the PMS and the young early-type stars in each 
subgroup of the Sco-Cen OB association and in the Chamaeleon region. It is 
remarkable that the two populations of young stars in this extended region do 
not show differences in terms of spatial distribution, space velocity distribution 
and age distribution. At least, in the US and UCL associations, for which we have 
more data, the young early-type stars and the PMS stars occupy approximately the 
same volume of space, have the same average space velocities and the same age, so 
that we must conclude that the two types of stars were born in the same event. 
The fact that the two types of stars are a single population has not always been 
recognized, probably for two reasons: 
1) the TT stars are difficult to detect in distant associations, and the term 
``OB association" seems to exclude the presence of stars of later spectral types; 
2) since low-mass stars take a longer time to reach the main-sequence than high-mass 
stars, they give the impression that they are a younger population.

It is interesting to note that a direct comparison of data in the literature could 
lead us to the incorrect conclusion that the dispersion of proper motions in the 
studied cluster is larger for the PMS stars than for the young early-type stars. 
This is the result of a selection effect, since the kinematic criterion used by dZ99 
to decide if an early-type star belongs to an OB association is very restrictive.

The Sco-Cen OB associations have slightly distinct properties, in term of ages and 
kinematics, that justify the division in three associations. 
Despite their differences in age and kinematics, these stellar associations bear 
sufficiently similar characteristics to suggest that they belong to a same large 
ensemble. Even after removal of the effect of the Solar motion, the OB subgroups 
show very ordered velocities in the plane of the sky. The differences between the 
space velocities are well inside the internal dispersion values of each association
(see Table 6).

We next discuss the gallery of models that have been put forward to explain at 
least some of the observed properties of the studied SFRs and the Sco-Cen OB
association.

\subsection{The sequential star-formation model}

The sequential star-formation model was mainly formulated by Blaauw (\cite{blaauw64}, 
\cite{blaauw91}), who considered that the US, UCL, and LCC associations were the 
prototype of such a mechanism. In this model, the star formation begins at one 
extremity of a giant molecular cloud, and propagates like a fire in a forest. 
The wind produced by the first group of stars compresses the gas of the cloud and 
induces the formation of a new group of stars, next to it. The star-formation 
history of the Sco-Cen OB association proposed by Preibisch \& Zinnecker 
(\cite{preib99}) is a variation of the sequential star-formation idea. 

Blaauw (\cite{blaauw64}, \cite{blaauw91}) did not make any prediction concerning 
proper motions, however we might expect to see the stars diverging with respect 
to the successive centers of star formation. The very ordered proper motion 
observed in the studied region does not seem to confirm this model. In addition, 
the absence of an age gradient, as has been shown in Sect. 5, excludes the 
existence of bursts of star formation at successive positions. The large spread 
of ages at any given position along the extended structure conflicts with the 
idea that the star-formation process propagated along it. 
Note that if we are convinced that sequential star formation was not the 
dominant star-formation mechanism in the region that was considered as the best
example of it, we can doubt its effectiveness in general.

\subsection{The Gould Belt model}

The expanding Gould Belt model was proposed by Lindblad et al. (\cite{lindblad}), 
Frogel \& Stothers (\cite{frogel}), and Olano (\cite{olano}), among others; 
a recent review was presented by P\"oppel (\cite{poppel}). The model supposes 
that a violent event took place about 30--60 Myr ago near the Cas-Tau group of 
young stars, giving rise to an expanding ring of gas. Olano (\cite{olano}) 
predicts the orientation of the ring, the distance to the Sun and velocity of 
the gas associated with it as a function of time. The Sco-Cen OB association would 
be a segment of that ring. For the best model, the distance (166 pc) of the 
portion of the ring situated in the third Galactic quadrant is in agreement with 
that of the Sco-Cen OB association. The model does not attempt to fit the space 
velocities of the young stars, but only the radial velocity of the H{\small I} 
gas. Anyway, from the model, we would expect gas and stellar motions with a 
component towards increasing longitudes, and not the contrary, as we observe. 
The range of ages of the stars that we observe is a large fraction
of the time elapsed since the starting event, according to Olano's model.
As the star-forming ring expanded, we would expect to see young stars filling
a large fraction of the volume inside the ring. There are, indeed, young stars
associated with the Taurus molecular cloud, but there is clearly a 
gap without young stars between the ``center" and the ``ring". In fact,
the distribution of young stars does not confirm the existence of a ring.
As later discussed, we consider that a segment a spiral arm is a better
explanation than a segment of ring.

\subsection{Star formation by impact of high-velocity clouds}

L\'epine \& Duvert (\cite{lepine94}) proposed star formation by impact of high 
velocity clouds (HVC) on the Galactic disk as a mechanism to explain the 
morphology of molecular clouds and age gradients of young stars; in particular, 
they presented a detailed model of the $\rho$ Oph cloud and Sco-Cen OB association. 
According to that model, the stars of the Sco-Cen OB association should have 
proper motions directed towards decreasing longitudes, reflecting the initial 
direction of the HVC, that can be inferred from the morphology of $\rho$ Oph and 
from the filaments of gas connecting the cloud to the Galactic plane. The model 
predicts a gradient of age in the same direction of space velocities. 
Although the model is able to explain the well-ordered proper motions, because 
the stars are expected to have similar initial velocities, a more detailed 
analysis, after correction of space velocities for the motion of the Sun, shows 
some difficulties. In particular, the model predicts a gradient of velocities 
(the first stars formed would have larger velocities and would be situated at 
a larger distance from the initial point of impact) and a gradient of age, that 
are not observed.
 
\subsection{Transient Cloud Cores}

Feigelson (\cite{feigel96}) analyzed the possibility of star formation in small 
transient cloud cores, that could be condensations in a turbulent giant molecular 
cloud. The author recognizes that there is an enormous flexibility in the range of 
parameters of the model. In general, the stars would be seen to diverge from local 
centers, and the younger TT stars would appear to be concentrated in clumps 
1--2$\degr$ in size. This model would not explain the ordered motion of the OB 
subgroups in Sco-Cen association. 
This model does not tell us what is the mechanism of molecular cloud formation or 
what drives the turbulence (in the two models previously discussed, the existence of 
molecular clouds is due to the compression of the interstellar medium by well-described 
shocks); the model could be complementary to a more global one, as described in the 
next section. 
 
\subsection{Purely random process}

We analyzed in some detail the possibility of purely random star formation.
Let us imagine that star-formation bursts occur at random places
and random times in the Galactic disk; they could be triggered for
instance by the infall of high velocity clouds or any other
process. We performed simple numerical experiments in order to estimate
the probability of obtaining by chance three stellar clusters close together
in a short interval of time, so that they woeld seem to be the result of  a
non-random mechanism. The experiments consisted of drawing circles
at random positions inside an area, with the correct scale to represent 
clusters in the Galactic plane. The typical diameter of recently
formed clusters is of the order of 50 pc, as suggested for instance by
Fig. 4. In a region of 1 kpc x 1 kpc around the Sun, in the
Galactic disk, about 10 clusters are formed in 10 Myr.
This time interval is such that if some of these clusters happen
to be close together, they would not appear to present any strong age 
gradient. The amount of clusters that we consider to be born in this 
time interval is about what we observe around the Sun, and
is not inconsistent with the star formation rate usually estimated
for the Galactic disk (about 1 to 10 stars/year, e.g., G\"usten \& Mezger
\cite{gusten}). The probability of obtaining three clusters close enough 
so that they form an elongated structure of length about 150 pc, even
without any restriction concerning the direction of the alignment, is smaller 
than 10\%. This points towards the need for a non-random triggering 
mechanism like a spiral arm, as described in the next section,
to produce an alignment of stellar clusters. Of course, if in addition, we 
considered the probability of obtaining aligned space velocities, 
the probability of obtaining such a structure would be much less.

\subsection{The spiral arm interpretation}

Finally, we must consider the possibility that what we are observing is a spiral 
arm that passes close to the Sun. Spiral arms are known to contain molecular 
clouds and OB stars, like the complex extending from Ophiuchus to Chamaeleon. 
The narrow alignment of molecular clouds in the Galactic plane revealed by CO 
(Dame et al. \cite{dame}) has precisely the inclination with respect to the Solar 
circle (about 15$\degr$) that we would expect from the global spiral structure 
(e.g., L\'epine \& Leroy \cite{lepine00}). Furthermore, the IRAS brightness profile 
of the Galaxy at 60 $\mu$m and at 100 $\mu$m as a function of longitude in the 
Galactic plane shows maxima in the directions $l \cong +82\degr$ and $l \cong -92\degr$ 
(Amores \& L\'epine \cite{amores}); these can be interpreted as tangential directions to 
an arm that passes close to the Sun. These peaks are similar to the other peaks 
associated with well-known tangential directions of spiral arms, like the features 
at $l = 30\degr$, $l = -56\degr$, etc. Note that a spiral arm at a Galactic radius 
larger than the Solar radius R$_{\odot}$ does not produce tangential directions; the 
directions of the observed peaks correspond to a spiral arm passing at a radius 
slightly less than R$_{\odot}$. Finally, the global distribution of H{\small II} 
regions in the Galactic plane also indicates the presence of this very close spiral 
arm inside the Solar radius, that produces concentrations of H{\small II} regions 
in the directions $l = 85\degr$ and $l = -95\degr$; this is discussed in detail by 
L\'epine et al. (\cite{lepine01}).

The question, then, is: what are the expected space velocities of young stars and 
molecular clouds recently formed in a spiral arm? Let us first comment briefly on 
motions in the direction perpendicular to the Galactic plane. A spiral arm is the 
site of gas shocks, and it would not be surprising if in such shocks, due to the 
increase in gas pressure, some gas were ejected to distances from the plane that are 
larger than the normal scale-height of the gas. This could explain the relatively 
high galactic latitudes of $\rho$ Oph and Chamaeleon clouds. Smoothed particle
magnetohydrodynamic simulations of gas cloud collisions show this effect of expansion 
of the gas in directions perpendicular to that of the collision (Marinho \& L\'epine 
\cite{marinho}).

Let us now consider the motion in the plane. The post-shock velocity of the gas, 
that is responsible for the initial velocity of the young stars, depends on the 
position of the arm relative to the corotation radius. We assume that the corotation 
radius lies beyond R$_{\odot}$, or very close to R$_{\odot}$ (Amaral \& L\'epine 
\cite{amaral}; Mishurov \& Zenina \cite{mishurov}). 
For radii smaller than the corotation radius, the stars 
and the gas rotate at velocities larger than that of the spiral pattern, so that 
the H{\small I} gas clouds orbiting in the Galactic plane reach the spiral arms 
on their concave side, and suffer a shock that strongly brakes and compresses 
the gas, leading to the formation of short-lived molecular clouds (seen as dust 
lanes in galaxies) and triggering star formation. The initial velocity of the formed 
stars is the same as their parent molecular clouds; since the clouds have 
just been braked, the young stars have a rotation velocity smaller than the normal 
circular velocity at that radius. As a consequence, the gravity force from the 
central parts of the Galaxy is no longer balanced by the centrifugal force; the young 
stars start an infall motion towards the Galactic center. The combination of this 
initial infall with the rotation velocity smaller than the normal rotation velocity 
is responsible for an initial trajectory that lies precisely along the spiral arm. 
This was already noticed by Bash (\cite{bash}) who calculated the ballistic orbits of 
H{\small II} regions and measured their infall. Outside the corotation radius, the 
contrary is expected: the gas is expected to be accelerated in the direction of 
circular motion by the shock.

So, inside the corotation radius, the youngest stars produced in spiral arms
should present velocities in the direction of the Galactic rotation, along the
arms, but with a velocity modulus smaller than that of normal rotation. Since
we are measuring the space velocities with respect to the LSR,  a frame that
rotates with the normal rotation velocity, we should see the young stars going in
the direction contrary to Galactic rotation, that is, in the direction of decreasing
longitudes, for objects in the third Galactic quadrant. This is precisely what we 
observe.

The spiral arm interpretation explains naturally why the velocity distribution is
uniform over an extended region, as well as the absence of a gradient of age. It 
is not surprising to find stars covering a large range of ages, since the arms 
are long lived structures, and due to their trajectories, recently formed stars 
are maintained in the arms for a relatively long time.

\section{Conclusions}

  \begin{itemize}
\item In the Chamaeleon region, there is an OB association, not previously 
catalogued, formed by at least 21 B- and A-type stars.

\item The extended group of young stellar associations, which includes US, 
UCL, LCC and Chamaeleon, form an almost continuous structure with a total 
length of at least 150 pc, with more uniform properties than previously 
believed. The young early-type stars of the OB associations and the PMS 
stars of the SFRs follow a similar spatial distribution, i.e., there is no 
separation between the low and the high-mass young stars. There are no 
differences in the kinematics or in the ages of the PMS stars and of the 
young early-type stars. 

\item The average space velocities of the studied young stars are uniform, 
directed contrary to Galactic rotation, towards decreasing galactic longitude. 

\item There is no measurable age gradient, all the sub-regions containing 
stars in the range of 1--20 Myr. Our results do not confirm that the UCL 
association is older than the LCC association. The absence of an age gradient 
excludes a star-formation process that propagates along the structure, like 
the traditional sequential star-formation process. 

\item We argue that the hypothesis that the Sco-Cen OB association is part 
of a spiral arm that passes close to the Sun explains most of the features 
of the space velocities and age distribution of the young stars. 
  \end{itemize}

\begin{acknowledgements}

The authors wish to thank Dr. R. de la Reza and Dr. C. A. O. Torres for 
helpful comments, and Dr. F. Adams for suggesting include a brief discussion
about random star formation. 
Special thanks are also extended to Dr. B. V. Castilho and Dr. J. I. B. Camargo.
MJS acknowledges the CAPES PhD fellowship, and FAPESP (No. 00/06954-6) and
CNPq (No. 300758/01-4) postdoc fellowships.
WSD acknowledges the FAPESP PhD fellowship No. 99/11781-4.
This work has been partially supported by PRONEX/Finep.
This research has made use of the SIMBAD database, operated at CDS, 
Strasbourg, France.

\end{acknowledgements}

\end{document}